\documentclass[12pt]{article}
\usepackage{amssymb,bm}
\usepackage{authblk}
\usepackage{amsmath,graphicx,color,epsfig}
\usepackage{slashed}
\usepackage{hhline}
\usepackage{setspace}
\usepackage{braket}

\author{M. Omar Nadeem}
\affil{\textit{Department of Physics, School of Natural Sciences, National University of
Sciences and Technology, H-12, Islamabad, Pakistan}}
\date{}

\title{Scalar Leptoquark Contribution to Two Photon Elastic Scattering}
\begin{document}
\maketitle

\begin{center}
\Large
\textbf{Abstract}
\end{center}
\normalsize
 We study the contribution of Scalar Leptoquark (SLQ) loops to the two photon scattering cross section at high energies. The leading order helicity amplitudes are discussed at coupling $\mathcal{O}(e^4)$ for Standard Model (SM) fermions, weak bosons and SLQs. Helicity amplitudes for the different types of loops are calculated for photon centre of mass energies being much higher than weak boson mass scale but still much lower than the predicted Leptoquark mass scales. Finally, the cross section due to the SLQ loops is presented in conjunction with the known SM contribution.
\normalsize

\section{INTRODUCTION}\label{sec:introduction}
Among the several candidates of new physics interactions, the notion of direct couplings of quarks and leptons has remained persistent throughout the years and is gaining renewed traction in recent times. While such phenomena are forbidden in the known SM, a certain class of new physics models do allow for such interactions by postulating the existence of bosonic particles called Leptoquarks which couple directly with quarks and leptons. This can be implemented either through an entirely different gauge group or an extension of the particle content of the known SM, or even a combination of the two.\\
The idea of Leptoquarks has been around since before 1974 when they appeared in the Pati-Salam Model~\cite{PSModel}, utilising the $SU(4)\times SU(2)_L\times SU(2)_R$ gauge group. In this paper we shall be considering an extension of the SM to include SLQs, as presented in~\cite{SLQfeynmanrules}. It postulates the existence of four types of massive scalar particles which carry both colour and electric charges. These Leptoquarks can couple up and down type quarks both among themselves and with leptons and their corresponding neutrinos. The bosonic sector of the Lagrangian also receives new interactions as Leptoquarks can couple to all of the SM bosons. This allows for a myriad of new topologies which have been used to explain anomalous phenomenon such as Lepton Flavour Universality violation~\cite{lfuv1,lfuv2,lfuv3} and the $W$ boson mass anomaly~\cite{wmass}.\\
This paper is aimed at indirect searches of SLQs by exploring the SLQ-Photon interactions in this model~\cite{SLQfeynmanrules} by analysing the SLQ loops' contribution to the cross section of elastic two-photon scattering at high energies. The $\gamma \gamma\rightarrow\gamma\gamma$ channel has been touted as useful tool to reveal new physics~\cite{twophotonnp} due to the relative ease of producing and controlling photons, as compared to $W$, $Z$ or Higg's bosons. The $\gamma \gamma\rightarrow\gamma\gamma$ channel also has a much cleaner background and fewer new parameters as compared to decay processes of heavier particles~\cite{twophotonnp}. As such this channel has been applied to the searches for Unparticles~\cite{unparticles} and investigating low scale quantum gravity~\cite{lsgravity,quantumgravity}.\\
Recent studies~\cite{slqmass} place the mass of Leptoquarks well into the TeV scale so to appreciably detect indirect leptoquark loop effects via the $\gamma\gamma\rightarrow\gamma\gamma$ channel or via $H\rightarrow gg/\gamma\gamma$ decays~\cite{De_2024} and even for direct searches by SLQ pair production~\cite{ghosh2023precise}, we require higher incident beam energies. This is becoming an increasingly attractive possibility due to the Photon Linear Collider (PLC)~\cite{plc} and LHC~\cite{lhc} being capable of achieving centre of mass energies well in excess of 200 GeV.\\
The layout of the rest of this paper is as follows: In Section 2 we present the kinematics of the process and in Section 3 we use it to calculate the leading helicity amplitudes in both SM and SLQ model. Section 4 presents the scattering cross section and in Section 5 we compare, via plots, the contributions of the various helicity amplitudes to the cross section and their angular distributions. Finally in Section 6 we give our conclusion and discuss the possible extensions to this work.\\
All Feynman diagrams in this paper have been made using the software\\JaxoDraw~\cite{JaxoDraw}.

\section{Kinematics}
The scattering process is considered in the centre of mass frame with the two initial photons colliding head on and the scattered photons going nearly collinear to the initial incident direction. Hence the cross section is analyzed at small scattering angle.\\
We shall be borrowing from the kinematics of~\cite{ibrarthesis} which presents the momenta and polarization vectors in the photons' centre of mass frame. The photon momenta are 
\begin{equation}\label{eq:1}
\begin{split}
k_{1}^{\mu}&=(E_{\gamma}, 0, 0, E_{\gamma})\;,\\
k_{2}^{\nu}&=(E_{\gamma}, 0, 0, -E_{\gamma})\;,\\
k_{3}^{\sigma}&=(E_{\gamma}, 0, E_{\gamma}\sin{\theta}, E_{\gamma}\cos{\theta})\;,\\
k_{4}^{\lambda}&=(E_{\gamma}, 0, -E_{\gamma}\sin{\theta}, -E_{\gamma}\cos{\theta})\;.
\end{split}
\end{equation}
Here we keep with the convention of associating a particular Lorentz index with each momentum four vector. The circularly polarized vectors for each photon are :
\begin{equation}\label{eq:2}
\begin{split}
\varepsilon_{\mu}^{1,\pm}&=\frac{1}{\sqrt{2}}(0,1,\pm i,0)\;,\\
\varepsilon_{\nu}^{2,\pm}&=\frac{1}{\sqrt{2}}(0,-1,\pm i,0)\;,\\
\varepsilon_{\sigma}^{3,\pm}&=\frac{1}{\sqrt{2}}(0,1,\pm i\cos{\theta},\mp i\sin{\theta})\;,\\
\varepsilon_{\lambda}^{4,\pm}&=\frac{1}{\sqrt{2}}(0,1,\mp i\cos{\theta},\pm i\sin{\theta})\;.
\end{split}
\end{equation}
Considering the energies at hand and the possible scale of Leptoquark masses, we do the calculations with a scale hierarchy. Leptoquark mass scale is much larger than photon energies hence the ratio $k_{i}.k_{j}/\hat{\mathbb{M}_{Q}^{2}}$ is considered a small parameter and its higher powers will result in the suppression of certain amplitudes and terms.
\section{Feynman Amplitudes}
The general form of all two photon scattering amplitudes is thus 
\begin{equation}\label{eq:3}
\mathcal{M}=\varepsilon^{*}_{\lambda}\varepsilon^{*}_{\sigma}\varepsilon_{\nu}\varepsilon_{\mu}\mathcal{M}^{\mu\nu\sigma\lambda}\;,
\end{equation}
with the Lorentz indices switched around for the different topologies. The rank four tensor structure $\mathcal{M}^{\mu\nu\sigma\lambda}$ shows up for all SM and SLQ contributions to the $\gamma\gamma\rightarrow\gamma\gamma$ amplitude. The leading order contribution $(e^4)$ includes all six flavours of the SM quarks and the three charged leptons which all form one loop box diagrams. Further contributions are from the $W^{\pm}$ bosons which give one loop bubble, triangle and box diagrams due to the $W^{\pm}\gamma$ triplet and quartic vertices. The four SLQs also give bubble, triangle and box one loop diagrams, as in Figure 1, with exactly the same coupling order as the SM particles' loops. Each amplitude will be written in helicity basis, determined by the helicity eigen state of each of the four participating photons.
\begin{equation}
\mathcal{M}^{r_1,r_2,r_3,r_4}=\mathcal{M}^{\mu\nu\sigma\lambda}\varepsilon_{\lambda}^{r_4*}\varepsilon_{\sigma}^{r_3*}\varepsilon_{\nu}^{r_2}\varepsilon_{\mu}^{r_1}\;,
\end{equation}\label{eq:4}
where $r_i=\pm$ denotes the helicity state of each photon. Due to parity invariance, time reversal symmetry and Bose symmetry , only some of the sixteen amplitudes are distinct~\cite{plc,ibrarthesis}. Those being $\mathcal{M}^{++++}$, $\mathcal{M}^{++--}$, $\mathcal{M}^{+-+-}$, $\mathcal{M}^{+--+}$ and $\mathcal{M}^{+++-}$.
\begin{figure}[htbp]
    \centering
    \includegraphics[width=0.9\textwidth]{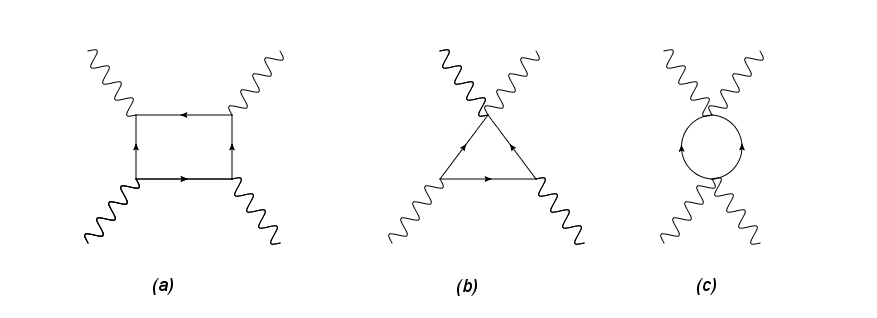}
    \caption{\textit{An example each of the (a) Box, (b) Triangle and (c) Bubble scalar Leptoquark loops in two photon scattering. Wiggly lines denote the photon and straight lines the Scalar Leptoquarks.}}
    \label{fig:SLQ loops}
\end{figure}

\subsection{Standard Model Amplitudes}
Even though our aim here is to investigate the effect of SLQ loops on the total cross section, SM amplitudes are essential to it because the observable cross section is proportional to $|\mathcal{M}|^2$ where $\mathcal{M}=\mathcal{M}_{SM}+\mathcal{M}_{SLQ}$.\\
The thesis~\cite{ibrarthesis} gives an excellent treatment of low energy fermionic contributions to the $\gamma\gamma\rightarrow\gamma\gamma$ cross section but it is only valid for photon energies much smaller than the light fermionic mass scales and hence unsuitable for our purposes where we consider energies much higher than $m_W$. As such the scale dependant fine structure is taken to be $\alpha=1/128$ and $m_w=80.22\text{GeV}$. Above this scale fermionic contributions are dominated by $W^{\pm}$ loop contributions~\cite{plc}. For photon energies much higher than the weak boson mass scale and at small scattering angles, taken to be $0^{\circ}<\theta<30^{\circ}$ the fermionic and bosonic leading helicity amplitudes are given in~\cite{plc} as
\begin{equation}\label{eq:5}
    \mathcal{M}_{W}^{++++}=\mathcal{M}_{W}^{+-+-}\;,
\end{equation}

\begin{equation}\label{eq:6}
    \mathcal{M}_{f}^{++++}=\mathcal{M}_{f}^{+-+-}\;,
\end{equation}

\begin{equation}\label{eq:7}
    \mathcal{M}_{W}^{++++}=16i\alpha^2\pi\frac{k_1.k_2}{k_1.k_3}\ln\left(\frac{k_1.k_2}{k_1.k_3}\right)\;,
\end{equation}

\begin{equation}\label{eq:8}
\mathcal{M}_{f}^{++++}=-4\alpha^2\ln^2\left(\frac{k_1.k_2}{k_1.k_3}\right)\;.
\end{equation}

\subsection{Scalar Leptoquark Amplitudes}

\begin{figure}[htbp]
    \centering
    \includegraphics[width=0.9\textwidth]{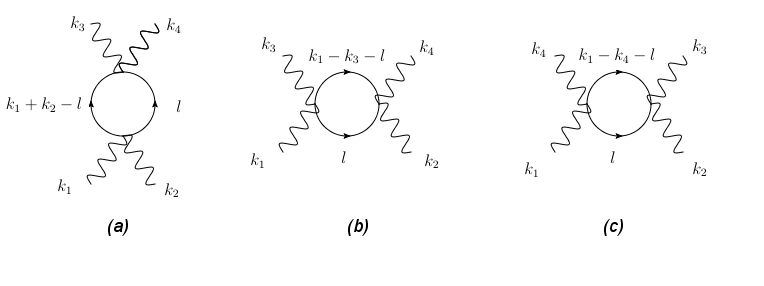}
    \caption{\textit{Each of these bubble diagrams has an equivalent accompanying diagram. This is because the loop consists of a positive SLQ and an equally negatively charged SLQ. In each case the accompanying diagram is due to the momenta swapped between the postive and negative SLQs.}}
    \label{fig:SLQ Bubbles}
\end{figure}
\noindent In keeping with the condition that the energy of the photons is much smaller than TeV scale ie $E_{\gamma}<<\hat{\mathbb{M}}_{Q}$, SLQ box loop contributions can be dropped as all the terms in those integrals scale as $1/\hat{\mathbb{M}}_{Q}^{4}$. SLQ triangle loops have a larger contribution but still smaller than SLQ bubble loops due to the significantly smaller pre-factor coefficients of the two triplet vertices~\cite{SLQfeynmanrules} in the triangle loop. Bubble loops are the ones giving the leading order contribution to the possible new physics in this process. Even so, the leading order term of the bubble loop is not interesting from a dynamical point of view as it has no dependance on photon energies and scattering angles. Those terms only show up in the subleading terms, scaling as $k_i.k_j/\hat{\mathbb{M}}_{Q}^{2}$.\\
Using the Feynman rules of~\cite{SLQfeynmanrules} one of the six bubble loops (Figure 2) is 
\begin{equation}\label{eq:9}
     \mathcal{M}_{B1}^{\mu\nu\sigma\lambda}=\sum_{X,Q}\mu^{2\epsilon}\int \frac{d^Dl}{(2\pi)^D}\frac{\text{Tr}[(ie^2\frac{X}{9})g^{\mu\nu}(ie^2\frac{X}{9})g^{\sigma\lambda}]}{[l^2-\hat{\mathbb{M}_{Q}^{2}}][(k_1+k_2-l)^2-\hat{\mathbb{M}_{Q}^{2}}]}\;.
\end{equation}
The sum is implied over the four different types of SLQs in this model with $Q$ being their electric charge and $X$ being a number in the vertex factor for each type of Leptoquark. Here $Q=\pm1/3,\pm2/3,\pm4/3,\pm5/3$ and $X=2,8,32,50$ respectively. Expanding the fully integrated result of \eqref{eq:9} in powers of the regulator $\epsilon$, using the package~\cite{hypexp}, gives

\begin{equation}\label{eq:10}
\begin{split}
\mathcal{M}_{B1}^{\mu\nu\sigma\lambda}&=\sum_{X,Q}-4i\alpha^2\left(\frac{X}{9}\right)^2g^{\mu\nu}g^{\sigma\lambda}\times\Biggl[\frac{1}{\epsilon}+1+\ln\left(\frac{\mu^2}{2\hat{\mathbb{M}_{Q}^{2}}}\right)\\
&+\frac{1}{2}\ln\left(1+\frac{k_1.k_2}{2\hat{\mathbb{M}_{Q}^{2}}}\right)-\sqrt{-\frac{2\hat{\mathbb{M}_{Q}^{2}}}{k_1.k_2}}\text{ArcTanh}\left(\sqrt{-\frac{k_1.k_2}{2\hat{\mathbb{M}_{Q}^{2}}}}\right)\Biggr]\;.
\end{split}
\end{equation}
This particular package properly works only for Mathematica versions 12.0 and higher~\cite{Wolfram12}. Another exactly equivalent loop, with the same tensor structure, is possible by swapping the loop momenta between the positive and negative SLQs in each loop. Hence
\begin{equation}\label{eq:11}
     \mathcal{M}_{B1}^{\mu\nu\sigma\lambda}= \mathcal{M}_{B2}^{\mu\nu\sigma\lambda}\;.
\end{equation}
Other loops are 
\begin{equation}\label{eq:12}
\begin{split}
\mathcal{M}_{B3}^{\mu\sigma\nu\lambda}&=\sum_{X,Q}-4i\alpha^2\left(\frac{X}{9}\right)^2g^{\mu\nu}g^{\sigma\lambda}\times\Biggl[\frac{1}{\epsilon}+1+\ln\left(\frac{\mu^2}{2\hat{\mathbb{M}_{Q}^{2}}}\right)\\
&+\frac{1}{2}\ln\left(1-\frac{k_1.k_3}{2\hat{\mathbb{M}_{Q}^{2}}}\right)-\sqrt{\frac{2\hat{\mathbb{M}_{Q}^{2}}}{k_1.k_3}}\text{ArcTanh}\left(\sqrt{\frac{k_1.k_3}{2\hat{\mathbb{M}_{Q}^{2}}}}\right)\Biggr]\;,
\end{split}
\end{equation}
\begin{equation}\label{eq:13}
    \mathcal{M}_{B3}^{\mu\sigma\nu\lambda}=\mathcal{M}_{B4}^{\mu\sigma\nu\lambda}\;,
\end{equation}
and
\begin{equation}\label{eq:14}
\mathcal{M}_{B5}^{\mu\lambda\nu\sigma}=\mathcal{M}_{B6}^{\mu\lambda\nu\sigma}\;.
\end{equation}
where \eqref{eq:13} is exactly the same as \eqref{eq:14} except for $k_1.k_3$ replaced with $k_1.k_4$. Hence, from the kinematics of the previous section, it is clear that \eqref{eq:9} only has photon energy dependance while \eqref{eq:10} and \eqref{eq:11} have dependance on the scattering angle as well. 
\\
The SLQ bubbles can be renormalized by invoking a renormalization condition that SLQ loop amplitudes are negligible at energies much smaller than electron mass scale and by expanding up to leading power in $x_{ij}=k_i.k_j/\hat{2\mathbb{M}_{Q}^{2}}$, the renormalized bubble loops become
\begin{equation}\label{eq:15}
    \mathcal{M}_{B1,R}^{\mu\nu\sigma\lambda}=\sum_{X,Q}-4i\alpha^2\left(\frac{X}{9}\right)^2g^{\mu\nu}g^{\sigma\lambda}\left[\frac{x_{12}}{6}\right]\;,
\end{equation}
\begin{equation}\label{eq:16}
    \mathcal{M}_{B3,R}^{\mu\sigma\nu\lambda}=\sum_{X,Q}-4i\alpha^2\left(\frac{X}{9}\right)^2g^{\mu\sigma}g^{\nu\lambda}\left[-\frac{5}{6}x_{13}\right]\;,
\end{equation}
\begin{equation}\label{eq:17}
    \mathcal{M}_{B5,R}^{\mu\lambda\nu\sigma}=\sum_{X,Q}-4i\alpha^2\left(\frac{X}{9}\right)^2g^{\mu\lambda}g^{\nu\sigma}\left[-\frac{5}{6}x_{14}\right]\;.
\end{equation}
Adding all the above bubble amplitudes gives
\begin{equation}\label{eq:18}
\begin{split}
    \mathcal{M}_{B,R}^{r_1,r_2,r_3,r_4}&=2(4i\alpha^2)\left(\sum_{X}\left(\frac{X}{9}\right)^2\right)\\
    &\times\Biggl[\left(-\frac{x_{12}}{6}\right)[(\varepsilon_{1}^{r_1}.\varepsilon_{2}^{r_2})(\varepsilon_{3}^{r_3*}.\varepsilon_{4}^{r_4*})]\\
    &+\left(\frac{5}{6}x_{13}\right)[(\varepsilon_{1}^{r_1}.\varepsilon_{3}^{r_3*})(\varepsilon_{2}^{r_2}.\varepsilon_{4}^{r_4*})]\\
    &+
    \left(\frac{5}{6}x_{14}\right)[(\varepsilon_{1}^{r_1}.\varepsilon_{4}^{r_4*})(\varepsilon_{2}^{r_2}.\varepsilon_{3}^{r_3*})]\Bigg]\quad.
\end{split}
\end{equation}

\section{Scattering Cross Section}
The unpolarized differential scattering cross section for $\gamma \gamma\rightarrow\gamma\gamma$ is given by~\cite{ibrarthesis}
\begin{equation}\label{eq:19}
    \frac{d\sigma}{d\Omega}=\frac{|\mathcal{M}|^2}{256\pi^2 E_{\gamma}^{2}}\;,
\end{equation}
where sum is implied over the two helicity states of each photon. To get the most significant term in the cross section, we combine the leading SM and SLQ contributions in each amplitude.
\begin{equation}\label{eq:20}
\begin{split}
\mathcal{M}^{++++}&=\mathcal{M}_{W}^{++++}+\mathcal{M}_{SLQ,B}^{++++}\;,\\
\mathcal{M}^{+-+-}&=\mathcal{M}_{W}^{+-+-}\;,\\
\mathcal{M}^{++--}&=\mathcal{M}_{SLQ,B}^{++--}\;.
\end{split}
\end{equation}
Hence
\begin{equation}\label{eq:21}
    |\mathcal{M}|^2=\frac{1}{4}(2|\mathcal{M}^{++++}|^2+2|\mathcal{M}^{++--}|^2+2|\mathcal{M}^{+-+-}|^2)\;.
\end{equation}
It can also be seen that the leading order new physics effects come from the cross terms of the total amplitude mod square, in the following form 
\begin{equation}\label{eq:22}
    |\mathcal{M}|^2_{W,SLQ}=\mathcal{M}_{W}^{++++}(\mathcal{M}_{SLQ,B}^{++++})^{*}+(\mathcal{M}_{W}^{++++})^{*}\mathcal{M}_{SLQ,B}^{++++}+...\;.
\end{equation}
The $W^{\pm}$ loop amplitude acts as the large coefficient for the much smaller SLQ amplitude.
\section{Numerical Results}
Figures 3 to 7 show the variation of the positive real values of the various helicity amplitudes contributing to this process. The general trend is that amplitudes increase upon increasing photon energies but the effect is tempered in the total differential cross section by the quadratic photon energy term in the denominator.\\
Figure 4 shows the variation of the magnitudes of the $W^{\pm}$ amplitudes with scattering angle, at fixed energy 300 GeV. It is seen to grow for larger scattering angles, a trend which is seen in the SLQ improved differential cross section as well, in Figure 9. Meanwhile in Figure 6, the leading SLQ bubble helicities decrease at higher scattering angles.\\
All the plots in Figures 3 to 9 have been made under the assumption that all four types of SLQs have equal masses of order 3 TeV. Only small corrections are noted by the addition of SLQ loops or by varying the predicted mass of SLQs. Nevertheless, such things do have an effect on the cross section, which is most evident by analyzing the cross section contribution from new physics effects as given in Figure 10. They provide an increase of the order $10^{-2}$ to the SM cross section. Figure 10 also shows the leading corrections to the cross section for three different cases of SLQ masses. It shows that higher values for the SLQ masses will result in smaller contributions to the cross section. By giving each SLQ a different mass, lying in the 2-8 TeV range, the new plot will lie between the 2 TeV and the 8 TeV plots.
\bigbreak
\begin{figure}[htbp]
    \centering
    \includegraphics[width=0.9\textwidth]{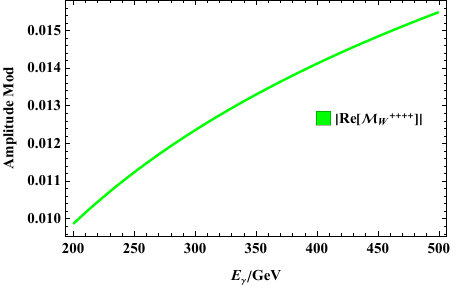}
    \caption{\textit{$W^{\pm}$ amplitudes dominate the differential cross section.}}
    \label{fig:Wenergy}
\end{figure}
\begin{figure}[htbp]
    \centering
    \includegraphics[width=0.9\textwidth]{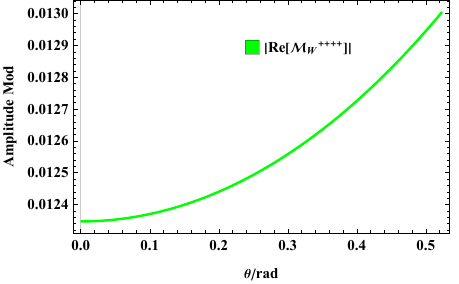}
    \caption{\textit{Angular behaviour of the $W^{\pm}$ loops' contribution at fixed energy.}}
    \label{fig:Wangular}
\end{figure}
\begin{figure}[htbp]
    \centering
    \includegraphics[width=0.9\textwidth]{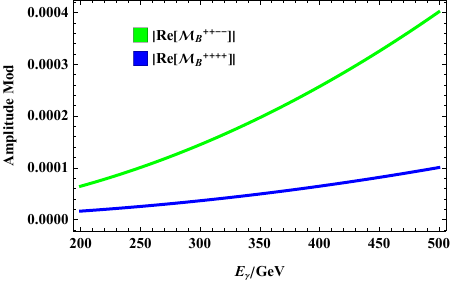}
    \caption{\textit{Leading new physics effects are due to two helicities of SLQ bubbles.}}
    \label{fig:BSLQenergy}
\end{figure}
\begin{figure}[htbp]
    \centering
    \includegraphics[width=0.9\textwidth]{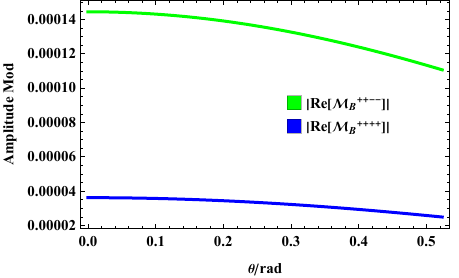}
    \caption{\textit{Angular dependance of the magnitude of the leading SLQ bubble helicities at fixed energy.}}
    \label{fig:BSLQangular}
\end{figure}
\begin{figure}[htbp]
    \centering
    \includegraphics[width=0.9\textwidth]{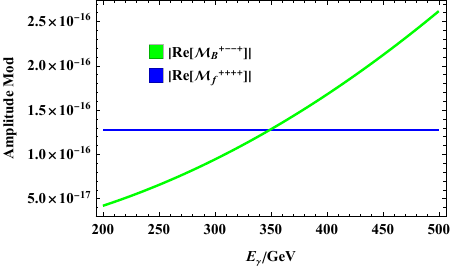}
    \caption{\textit{Suppressed helicities of bubbles and fermionic loops.}}
    \label{fig:{Suppressed}}
\end{figure}
\begin{figure}[htbp]
    \centering
    \includegraphics[width=1.0\textwidth]{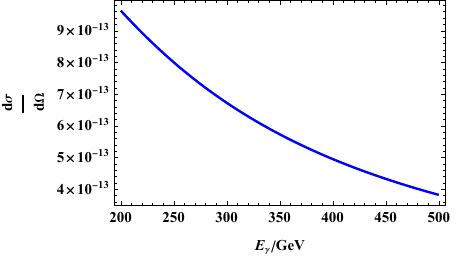}
    \caption{\textit{The profile of the SLQ model improved cross section closely follows the one due to the SM, as the overall magnitude is set by the $W^{\pm}$ contribution.}}
    \label{fig:TCSe}
\end{figure}
\begin{figure}[htbp]
    \centering
    \includegraphics[width=1.0\textwidth]{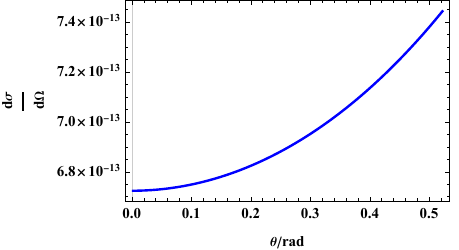}
    \caption{\textit{Angular behaviour of the SLQ improved differential cross section. This also closely follows the one due to SM alone.}}
    \label{fig:TCSa}
\end{figure}
\begin{figure}[htbp]
    \centering
    \includegraphics[width=1.0\textwidth]{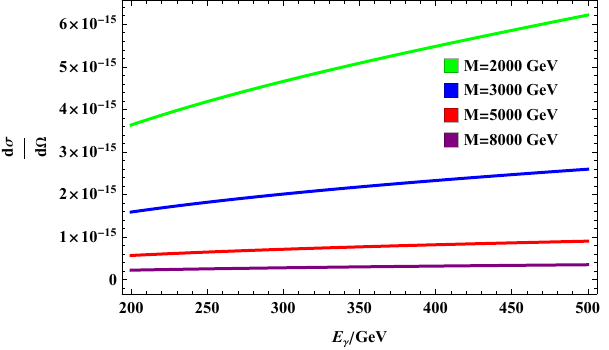}
    \caption{\textit{Cross section contributions from different values of SLQ mass.}}
    \label{fig:NP LO terms Diff Cross Section}
\end{figure}
\pagebreak
\section{Conclusion}
For energies in the high GeV scale, the contribution of SLQ one loop amplitudes to the two photon scattering cross section is smaller than the $W^{\pm}$ contribution but still significantly larger than that of SM fermionic loops. The SLQ improved cross section closely follows the trend of the SM predictions with regards to variations with respect to both energy and scattering angle. This is due to the fact that the tone of the overall process is being set by the significantly larger $W^{\pm}$ loops which dominate over the effects of $\mathcal{M}_{SLQ,B}^{++--}$ and $\mathcal{M}_{SLQ,B}^{++++}$. Other SLQ bubble helicities are comparable to the SLQ triangle helicities and fermionic loops, being several orders of magnitude smaller than the leading terms. Higher predicted values of the SLQ mass will result in progressively smaller cross section contributions.\\
These results can be significantly improved by QCD corrections to the SLQ loops. While QCD corrections have been applied to SM fermionic loops~\cite{2loopLightbyLight}, at these energies even the leading order one loop fermionic amplitude is significantly suppressed. The $W$ boson loops on the other hand, only allow for the much smaller QED corrections. As such, the QCD improved SLQ loops could come very close to the QED improved $W$ boson loops.

\pagebreak

\section*{ACKNOWLEDGEMENTS}
I extend my gratitude to my colleagues I.Hussain,B.Altaf and F.Ahmad for vital tech support. And to F.M.Bhutta and S.Ishaq for several helpful discussions and feedback.

\bibliographystyle{ieeetr}
\bibliography{bibliography.bib}
\end{document}